\documentstyle[aps,preprint]{revtex}
\input amssym.def
\input amssym.tex
\newcommand{\lax}{La_{1-x}A_xMnO_3}
\everymath{\rm}
\everydisplay{\rm}
\begin{document}
\title
{Cooperative Jahn-Teller Effect and Electron-Phonon Coupling
in ${\bf La_{1-x}A_xMnO_3}$}
\author{A. J. Millis}
\address{AT\&T Bell Laboratories \\
600 Mountain Avenue  \\
Murray Hill, NJ 07974 }
\maketitle
\renewcommand{\baselinestretch}{2}
\begin{abstract}
A classical model
for the lattice distortions of $\lax$
is derived and, in a mean field approximation, solved.
The model is based on previous work by Kanamori
and involves
localized Mn d-electrons
(which induce tetragonal distortions of the oxygen
octahedra surrounding the Mn) and localized holes
(which induce breathing distortions).
Parameters are determined by fitting to
the room temperature structure of
$LaMnO_3$.
The energy gained by formation of a local lattice distortion
is found to be large, most likely
$\approx 0.6$ eV per site, implying a strong electorn-phonon
coupling and supporting polaronic models of transport
in the doped materials. The structural transition
is shown to be of the order-disorder type;  the
rapid x-dependence of the transition temperature
is argued to occur because added holes
produce a "random" field which misaligns the nearby sites.
\end{abstract}
\newpage

$LaMnO_3$ is an insulator which undergoes a structural phase
stransition at a $T_s (x=0) \sim 750K$.
The high temperature phase is believed
to be cubic.
The low temperature phase is approximately
tetragonal, with one lattice constant
rather shorter than the other two\cite{Wollan55}.
Several other rather small amplitude distortions
also occur at temperatures less than or equal
to $T_s$\cite{Ellemans71},
and the structure at room temperature is
orthorhombic.
These small distortions will be ignored here.
As the composition is varied to $\lax$,
there are two changes.
First, $T_s(x)$ decreases rapidly and vanishes
at $x = x_s \approx 0.2$\cite{Wollan55,Ellemans71}.
Second, the resistivity decreases\cite{Tokura94}.
However, for $x < x_{cond} \approx 0.3$ and temperatures
of order room temperature and higher, the material
is still insulating in the sense that the resistivity
is much higher than the Mott limit, and increases
at T is decreased \cite{Millis95a}.
In this regime a description of the resistivity
in terms of classical particles hopping on a
lattice has been shown to be self consistent\cite{Millis95b}.

This paper presents a model for the $x < x_{cond}$ regime and
an explanation for the x-dependence of $T_s$.
The physical picture is as follows:
the electrically active orbitals are believed to be
the Mn $d_{3z^2-r^2}$ and $d_{x^2-y^2}$ orbitals.
The mean occupancy is $1-x$\cite{Wollan55,Goodenough55}.
Because the conductivity is so low, the electrons are treated
classically.  It is assumed that a site is occupied, with
probability $1-x$ or empty, with probability $x$.
The  d-orbitals are degenerate if the local
environment has cubic symmetry;
the degeneracy is lifted by a tetragonal distortion
of the local environment.
Kanamori\cite{Kanamori59} deduced
that at $x=0$ the primary lattice distortion
occurring at $T_s$ is a staggered
($\pi , \pi , \pi$) tetragonal distortion
of the oxygen octahedra surrounding the Mn sites, driven
by a Jahn-Teller splitting of the outer Mn d-levels;
anharmonic terms in the elastic energy couple
this to the uniform strain, producing the lattice parameter
changes observed in early scattering experiments.
Kanamori's deduction was subsequently confirmed
by more detailed studies of the structure\cite{Ellemans71}.
In this paper ionic displacements will be
explicitly included in Kanamori's model,
a fit to data will be given, and the model will be
extended to $x>0$.
It will be shown that the energies involved in the
Jahn-Teller physics are much larger than any relevant
temperature, so that as long
as a classical picture for the electrons is appropriate,
a local tetragonal distortion will occur around each Mn
site where there is an outer shell electron.
At each unoccupied site a breathing mode distortion
will occur; this will act as an effective random
field on the staggered tetragonal distortions,
and will prevent them from ordering.
If the tetragonal distortions are not coherent throughout
the lattice, they cannot couple to the uniform
strain, and the material will remain approximately cubic.

The model considered here is a version of the
``cooperative Jahn-Teller effect'',
which has generated an enormous literature \cite{Kugel82}.
Surprisingly, rather little attention has been paid to $LaMnO_3$
since the pioneering work of Kanamori.  A Hamiltonian describing
the orbital ordering of $LaMnO_3$
was derived from a purely electronic multiband
Hubbard model by Kugel and Khomskii \cite{Kugel73}
and a similar Hamiltonian has
recently been derived and studied via mean field theory by Ishihara
et. al. \cite{Ishihara95}, but atomic displacements and
electron-phonon coupling have not been considered.
The new aspects of the present paper are the explicit inclusion
of the lattice degrees of freedom, which allows values
for the electron-phonon coupling in $LaMnO_3$ to be
deduced from data, and the discussion of the ``random field''
effect of holes.

The rest of this paper is organized as follows.
In Section II the model is derived.
In Section III the parameters are determined
by fitting structural data for
$LaMnO_3$ to the model.
In Section IV the effects of added holes are discussed.
Section V is a conclusion.
Technical details of calculations are given in several Appendices.
\setcounter{section}{1}
\section{Model}
In this section the energy functional is derived.
The main physical assumption is that all degrees
of freedom may be treated classically.
The electrons are regarded as the fundamental
degrees of freedom and are taken to be localized on lattice
sites.  In a classical model the hopping of electrons from
site to site
does not affect the energy, so will be neglected. Note also that
the physical mechanism primarily responsible for localization
could be the electron-phonon coupling discussed here or
the "Hubbard-U" effects considered by other authors \cite{Inoue95}.
The cause of the localization is not relevant to the considerations
of this paper, so the electronic correlation effects need
not be explicitly considered.

For each fixed configuration of electrons, the phonon
part of the free energy is minimized; the result
of this minimization is the energy of that
configuration of electrons.
The phonons are treated in the harmonic approximation.
The effect of the undoubtedly important anharmonic
terms in the lattice energy is parametrized.
Only some of the lattice degrees of freedom are considered.
These are: 1) the vector displacements
$\vec{\delta}_i$ of the manganese (Mn) ion on
site $i$ and 2) the scalar displacement $u_i^{(a)}$ of
the oxygen (O) ion along the Mn-O-Mn bond direction.
Thus $u_i^x$ is the displacement, in the x direction,
of the O atom which sits between the Mn ion on site $i$
and the Mn ion on site $i+ \hat{x}$.
With this restricted set of displacements
one may discuss the Jahn-Teller distortion and the
uniform strain, but not the buckling of the Mn-O-Mn
bond or the associated rotation of the octahedra.
These latter lattice distortions occur but,
I believe, are not fundamental.

If an electron is present on site $i$, it will
be in a state $| \psi_i ( \theta ) >$ given by a
linear combination of the two outer d-orbitals.
In the classical approximation used here the phase
of the electron is of no significance, so one may write
\begin{equation}
| \psi_i ( \theta_i) > =
cos \theta_i | d_{3z^2-r^2} > +
sin \theta_i | d_{x^2-y^2} >
\label{eq:phi_i(theta )}
\end{equation}
with
$0 \leq \theta_i < \pi$.

The lattice energy $E_{Latt}$ is taken  to depend on the Mn-O
distance and the Mn-Mn distance.
The unit cell $i$ is taken to include the Mn ion
at position $\vec{R}_i + b \vec{\delta}_i$  and
the three O-ions at positions
$\vec{R}_i + \left ( \frac{1}{2} b+ b u_i^{(a)} \right ) \hat{a}$,
where $a=x,y$, or $z$ and $b$ is the lattice constant.
Here $\delta$ and $u$ are defined with reference to the ideal
perovskite lattice with lattice constant $b$.
In the harmonic approximation
\begin{equation}
\begin{array}{rl}
E_{Latt} & = \frac{1}{2} K_1 \sum_i
( \delta_i^a - u_i^a)^2 +
( \delta_i^a-u_{i-\hat{a}}^a )^2 \\
  & + \frac{1}{2} K_2  \sum_i
( \delta_i^a - \delta_{i-\hat{a}}^a )^2
\label{eq:E_{Latt}}
\end{array}
\end{equation}
Here $K_1$ and $K_2$ have the dimension of energy;
$\delta$ and $u$ are dimensionless.  One expects
$K_1 \ge K_2$.

If an electron is present on site $i$,
there is an electron-lattice energy given by
\begin{equation}
E_{JT} = \lambda \sum_i (1-h_i)
[ cos 2 \theta_i [v_i^z - \frac{1}{2} (v_i^x+v_i^y) ]
+ sin 2 \theta_i \frac{\sqrt{3}}{2}
[v_i^x -v_i^y ]]
\label{eq:E_{JT}}
\end{equation}
Here
\begin{equation}
v_i^a = u_i^a -u_{i-\hat{a}}^a
\label{eq:v_i^a}
\end{equation}
and $h_i = 0$ if an electron is present on
site $i$ and $h_i=1$ if not.
Finally, if there is no electron present
on site $i$, all of the neighboring oxygen
ions are equally attracted to it leading to
\begin{equation}
E_{hole} = \beta \lambda \sum_i h_i
[ v_i^x + v_i^y+v_i^z ]
\label{eq:E_{hole}1}
\end{equation}
One expects $\beta \gg 1$ because the force exerted
on the surrounding oxygen ions by a Mn of the wrong
charge must be much greater than the
force exerted by rearranging the proper charge
among different d-orbitals.

For fixed values of $\theta_i$ and $h_i$,
Eqs.  \ref{eq:E_{Latt}},\ref{eq:E_{JT}}, \ref{eq:E_{hole}1}
may be minimized.  The details are given in Appendix A.
The result is most naturally expressed in terms of
the parameters
\begin{equation}
\begin{array}{rl}
E^0 & = - \frac{3}{2} \frac{\lambda^2}{K_1}
\frac{K_1+K_2}{K_1+2K_2} \\
\label{eq:E^0}
\kappa & = \frac{1}{2} \frac{\lambda^2K_2}{K_1(K_1+2K_2)}
\label{eq:kappa}
\end{array}
\end{equation}
as
\begin{equation}
\begin{array}{rl}
E &= E^0 \sum_i (1-h_i)^2 + \beta^2 h_i^2 + A \; cos 6 \theta_i \\
   &+ \kappa \sum_{ia} (1-h_i) (1-h_{i+\hat{a}})
cos^2 ( \theta_i+ \psi_a) cos^2 ( \theta_{i+a} + \psi_a) \\
  &+ 2 \beta \kappa \sum_{ia} h_i (1-h_{i+a})
cos 2 ( \theta_{i+d_i} + \psi_a) +
\beta^2 \kappa \sum_{ia} h_i h_{i+a}
\label{eq:E}
\end{array}
\end{equation}
Here $a = \pm x,y,z$,
$\psi_{\pm z} = 0$,
$\psi_{\pm x} = - \pi / 3$,
$\psi_{\pm y} = \pi /3$, and we have
followed Kanamori \cite{Kanamori59}
by adding
a phenomenological anharmonicity term with coefficient A.
Cubic anharmonicities exist in any realistic
model of lattice dynamics.
The anharmonicity is important for two reasons:
it couples a staggered distortion to a uniform one,
and it breaks the
perfect rotational ($\theta$)
symmetry found otherwise if $h_i=0$.
The term added to Eq. \ref{eq:E} is the simplest
one which accomplishes there two effects and goes
into itself under $\theta \rightarrow \theta + \pi$
as required.
It is derived in Appendix A.

To each configuration of orbital occupancies
$\{ \theta_i \}$ corresponds an average distortion
from the ideal cubic peroviskite structure.
This may be written
in terms of the oxygen (u) and Mn ($\delta$) displacements as
\begin{equation}
\begin{array}{rl}
u_i^a & = \sum_{j} \phi_u^a (R_i-R_{j})
[(1-h_{j} ) cos (2 \theta_{j} + \psi_{j}^a) +
\beta h_{j} ] \\
\label{eq:ur}
\delta_i^a & = \sum_{j} \phi_{\delta}^a
(R_i-R_{j})
[(1-h_{j}) cos (2 \theta_{j} +
\psi_{j}^a) + \beta h_{j} ]
\label{eq:deltar}
\end{array}
\end{equation}
The elastic kernels are:
\begin{equation}
\begin{array}{rl}
\phi_u^a (R) & =
\frac{\lambda}{K_1} \sum_k
\frac{e^{ik \cdot R} (1-e^{-ik_a} )
 (K_1+K_2(1- cos k_a))}{(K_1+2K_2) (1- cos k_a)} \\
\label{eq:phi_u}
\phi_{\delta}^9 (k) & =
\frac{\lambda}{K_1} \sum_k
\frac{e^{ik \cdot R} K_1 cos (k_a /2)
(1-e^{-ik_a} )}{K_1+K_2 (1-cos k_a)}
\label{eq:phi_{delta}}
\end{array}
\end{equation}
\section{Fit to Data}
In this section the structural information
of\cite{Ellemans71} is used to estimate
model parameters.
The analysis is essentially that of Kanamori \cite{Kanamori59}.
A two-sublattice ordering of Jahn-Teller
distortions parametrized by angles $\theta_1$ and
$\theta_2$ is assumed.
By fitting the observed atomic displacements to
Eqs. \ref{eq:bar{delta}} and \ref{eq:baru}, $\theta_1$,
$\theta_2$ and elastic constants are determined.
By requiring that the deduced
$\theta_1$, $\theta_2$ minimize
Eq. \ref{eq:E}  the anisotropy energy A is found.
The experimental data for the structure are given in
Appendix B and the mean field equations are solved in
Appendix C.

It is convenient to express the lattice distortions
in terms of a staggered oxygen displacement
$\vec{u}_s$ and a uniform strain $\vec{e}$.
By rewriting Eqs.  \ref{eq:bar{delta}}, \ref{eq:baru}
we obtain
\begin{equation}
\begin{array}{rl}
u_s^a & = \frac{\lambda}{2K_1}
( cos 2 ( \theta_1 + \psi^2 ) - cos
2(\theta_2 + \psi^a)) \\
\label{eq:u_s}
e^a & = \frac{2 \lambda}{(K_1+2K_2)}
( cos (2 (\theta_1+\psi^a)) +
cos 2 (\theta_2 + \psi^a)))
\label{eq:e}
\end{array}
\end{equation}

In Appendix B the values
$e^a = - .028 (-1/2 , -1/2, 1)$
and $u_s^a=0.038(1, -1, 0)$ are derived
from the data of Ref \cite{Ellemans71}.
That $u_s^z =0$ implies
$\theta_2 =- \theta_1+\pi$; substituting this into
Eqs. \ref{eq:u_s} leads
to equations for $\theta_1$ and $\lambda / K_1$
which may be solved if $K_2 /K_1$ is given.
Results are listed in Table 1.

We now turn to the value of A.
The assumption of a two sublattice
distortion and the condition $h_i=0$ implies
that Eq. \ref{eq:E} becomes
\begin{equation}
E= \frac{1}{2} A [ cos (6 \theta_1) +
cos (6 \theta_2)] +
3 \kappa cos ( 2 \theta_1 - 2 \theta_2)
\label{eq:E2}
\end{equation}
By minimizing Eq. \ref{eq:E2} and using
$\theta_2 = \theta_1 + \pi$ we find
\begin{equation}
\frac{A}{\kappa} =-
\frac{-2 sin (4 \theta_1 )}{sin 6 \theta_1}
\label{eq:A}
\end{equation}
Values for $A/ \kappa$ are also listed in Table 1.

The most important information contained in Table I
is that the basic Jahn-Teller energy $E_0$ is
much greater than the stiffness $\kappa$ which
orients the distortions from site to site.
Indeed, from Eq. \ref{eq:E^0} the ratio may be seen
to be $\frac{1}{3}K_2/(K_1+K_2)$; as it is unlikely
that the Mn-Mn force constant $K_2$ $>$ the Mn-O force
constant $K_1$, the ratio is less than $1/6$.
The structural transition occurring at
$T_s \approx 800K$ in $LaMnO_3$ is therefore
of the order-disorder type, and we may expect local
distortions to persist for $T > T_s$.

{}From Table I it is also clear that the
anisotropy energy is not small, although
the precise value depends sensitively on $K_2 / K_1$.

Now consider magnitudes of energy scales.
The basic scale is $K_1$; this is related
to the frequency of an oxygen bond stretching
phonon $\omega_{ox}$ by
\begin{equation}
\omega_{ox} =
\sqrt{\frac{2K_1 /b^2 \hbar^2}{M_{ox}}}
\label{eq:omega_{ox}}
\end{equation}
The factor of two arises because there are two
Mn-O bonds in Eq. \ref{eq:E_{Latt}}.
Estimating $100$ meV$\gtrsim \hbar \omega_{ox} \gtrsim 30$ meV and
using $b=4 \AA$ gives
\begin{equation}
300 eV \gtrsim K_1 \gtrsim 30 eV
\label{eq:K_1}
\end{equation}
I am unaware of measurements of the phonon spectrum in
$LaMnO_3$.  If, however, it is assumed that
the phonon spectrum has a rather weak doping dependence
one may use data optical from $La_{1.85}Sr_{0.15}MnO_3$
\cite{Okimoto95}.
The highest-lying phonon modes were observed at
$\omega_{ph} \sim 70$
meV.  It is reasonable to assume that these are
the bond-stretching
oxygen modes of interest and that these modes are
only weakly dispersive; thus, one may identify
$\omega_{ph}$ with $\omega_{oxy}$ and estimate $K_1 \approx
200$ eV.

An alternative estimate may be obtained from the
mean-field approximation to the structural transition
temperature $T_s \approx 750K$.
This is shown in Appendix E to be
$T_s^{mF} \approx 3 \kappa$, and mean-field theory
overestimates $T_s$, so
\begin{equation}
\kappa > 20 meV
\label{eq:kappa>}
\end{equation}

This bound on $\kappa$ yields $K_2/K_1$-dependent
bounds for $K_1$ ranging from $K_1 > 220$ eV
($K_2 / K = 0.1$) to $K_1 > 50$ eV
($K_2 / K_1 =1$).
Values of $K_2 / K_1 \gtrsim 0.5$ are most consistent
with estimates of $\omega_{oxy} \lesssim 50$ meV;
those of $K_2 / K_1 < 0.5$ with
$\omega_{oxy} \gtrsim 50$ meV.  Combining this with the
estimate $K_1 \approx 200$ eV suggests $A \sim \kappa$.
This estimate is consistent with estimates given
in a standard review \cite{Sturge67} that typical anharmonicity
energies are of order a few hundred kelvin.

The estimates of $K_1$ imply Jahn-Teller energies
$E_0$ ranging from $\approx 100$ meV at the low end
($K_1 \sim 30$ eV)
to 1 eV at the high end ($K_1 \sim 300$ eV).
The estimate $\omega_{oxy} = 70$ meV implies
$E_0 \approx 0.6$ eV, slightly larger than the largest
Jahn-Teller energy listed in a standard review
\cite{Sturge67}.  In any event, because
the energy splitting between the two d-levels
is $4E_0$, it is safe to assume that at any
reasonable temperature the splitting is frozen
in.
Unfortunately the splitting is difficult to
measure directly because most methods for coupling
to the d-level involve changing the valence of the Mn,
which would bring other physics in to play.
The transition should be Raman active, though.

To summarize, it has been shown in this section that the Jahn-Teller
energy of $LaMnO_3$ may be written
\begin{equation}
\begin{array}{rl}
E_{x=0} &= \kappa \sum_{ia} cos (2 \theta_i + 2 \psi_a )
cos (2 \theta_{i+a} + 2 \psi_a) \\
     &+ A \sum_i cos 6 \theta_i
\label{eq:E_{x=0}}
\end{array}
\end{equation}

If $A \gtrsim T_s$
then it is reasonable to assume that at each
site $\theta_i$ is near one of the three angles
favored by the anharmonicity term, so that
the system may be mapped on to a three-state Potts
model as previously noted \cite{Kugel82}.
Details are given in Appendix D.
The result is conveniently written in a notation in which
the state of site $i$ is represented by a vector
$Q_i$ with a 1 in one place and 0 in the other two places;
$Q_i=(1,0,0)$ implies there is a Jahn-Teller distortion
with long axis along x, $Q_i=(0,1,0)$ means $y$
and (0,0,1), $z$.
Then
\begin{equation}
E_{Potts} = \kappa \sum_{ia} \vec{Q}_i
{\bf I}^a \vec{Q}_{i+c}
+ J^{\prime} \sum_{ia} \vec{Q}_i \cdot
\vec{Q}_{i+2a}
\label{eq:E_{Potts}}
\end{equation}
with ${\bf I}^a$ a bond-direction
dependent interaction given in Eq. \ref{eq:I}
and $J^{\prime} < 0$ a ferromagnetic
interaction between ``straight-line''
second neighbors which is of order $\kappa/A$
and was apparently not neglected in
previous work.
The second neighbor interaction is an
approximation to the true interaction, as discussed in
Appendix D.
The three state Potts model has a transition
in the x-y universality class as, therefore,
does Eq. \ref{eq:E_{x=0}}.
The second neighbor ``ferromagnetic''
coupling lifts the degeneracies which lead
to subtleties in the behavior of the
usual Potts model.  The estimates of A suggest that the extreme Potts
limit will not provide a good quantitative description of $LaMnO_3$.
\section{Holes}
This section discusses the effects of
added holes.
It is clear from Eq. \ref{eq:E} that a
hole on site $i$ eliminates the Jahn-Teller
distortion on site $i$ and leads to a potential,
$\beta \kappa cos (2 \theta_{i+b} + 2 \psi_b)$,
which acts to orient the distortion on site
$i+b$ so that its long axis is along $\hat{b}$.
Thus added holes lead both to site dilution and to a field
which tends to orient some of the neighbors of the hole
in directions not compatible with long range order.

If $A>0$ (as seems to occur in $LaMnO_3$) the angles favored
by holes are
compatible with the angles favored by
anharmonicity; if $A<0$ an interesting
competition arises, which will not be discussed here.

In the $A \gg 0$ limit the effect of added
holes is particularly transparent.
By following the derivation that led to
Eq. \ref{eq:E_{Potts}} one finds that a hole on
site $i$ produces a term in the energy
\begin{equation}
E_{hole}^i = \beta \kappa
\sum_b \vec{R}^b \cdot \vec{Q}_{i+b}
\label{eq:E_{hole}}
\end{equation}
with $R^x = (-1, 1/2, 1/2)$ etc. Thus in this limit
a hole manifestly produces a field which tends to orient
the spins on
neighboring sites.

A Monte Carlo investigation based on
Eq. \ref{eq:E} or on Eqs. \ref{eq:E_{Potts}},
\ref{eq:E_{hole}} would be desirable.
Here simple arguments are given to estimate $T_s(x)$.
Assume the hole positions are uncorrelated
with each other or with the configuration
of Jahn-Teller orderings.
To estimate the critical concentration, $x_c$,
of holes at which ordering vanishes, note that
for site diluted systems ($\beta =0$),
$T_s$ vanishes when the occupied sites
do not percolate \cite{Ziman89}.
For the simple cubic lattice, the percolation
threshold is about $p_c=0.3$ \cite{Ziman89}
so $x_c( \beta =0)=0.7$.
Of course for such large values of $x$
the model is not valid.
For $\beta \rightarrow \infty$, each hole eliminates 5
sites (itself and 4 neighbors:
two remain approximately correctly oriented),
implying $1-5x_c=.3$ or $x_c( \beta \rightarrow \infty ) \cong 0.14$.

Alternatively, one may use mean field theory
to estimate $T_s(x)$.
The fundamental object in mean field theory
is the probability distribution $P( \theta )$
of the angle on a distinguished site
in an effective field depending on the average
values of the angles on the adjacent sites
and on whether or not holes are present.
The assumption of uncorrelated holes implies
\begin{equation}
P( \theta ) = \sum_{\{h_a\}}
\frac{e^{-E( \theta , \{ h_a\} )/T}}{Z( \{ h_a \} )}
x^{n_h} (1-x)^{6-n_h}
\label{eq:P(theta)}
\end{equation}
Here $\{ h_a \}$ is a distribution of holes
on sites adjacent to the distinguished one,
$n_h$ is the number of holes in that particular
configuration, and
\begin{equation}
Z( \{ h_a \} ) = \int_0^{\pi}
\frac{d\theta}{\pi}
e^{-E(\theta , \{ h_a \} )/ T}
\label{eq:Z}
\end{equation}

The energy may be written in terms of the average
values of the cosine and sine on the other sublattice,
$c= \langle cos 2 \theta \rangle$ and
$s= \langle sin 2 \theta \rangle$ as
\begin{equation}
\begin{array}{rl}
E( \theta , & \{ h_a \} ) 2 \kappa
\sum_a (1-h_a) cos (2 \theta +2 \psi_c) \\
   & \cdot [(c \; cos 2 \psi_a -s \; sin 2 \psi_c )
    1-h_a + \beta h_c ]
\label{eq:E(theta)}
\end{array}
\end{equation}
The quantities $c$ and $s$ satisfy a self-consistency
equation; the linearized equation giving $T_s$ may
be written
\begin{equation}
c=- \int_0^{\pi} \frac{d\theta}{\pi} \;
P ( \theta ) \; cos 2 \theta
\label{eq:T_s-1}
\end{equation}

The derivation and evaluation of this equation
are given in Appendix E.
An analytic treatment is not simple
except in the limits $A \rightarrow 0$ or
$A \rightarrow \infty$ (arbitrary $\beta$)
and $\beta \rightarrow 0$ or $\beta \rightarrow \infty$
(arbitrary A).
For $A=0$,
\begin{equation}
1 = \frac{3 \kappa}{T_s}
\left [ 1-x \left ( 1+ \frac{4I_1^2}{I_0^2} +
\frac{I_2}{I_0} \right ) \right ]
\label{eq:T_s-2}
\end{equation}
Here the $I_n$ are Bessel functions of imaginary argument
$i \beta/T_s$.

In the $A \rightarrow \infty$
limit, $T_x(\beta , x)$ satisfies
\begin{equation}
1= \frac{3\kappa}{T_s}
\left [ 1-x \frac{6-3e^{-3\beta / 2T_s}+6e^{-3\beta /T_s}}
{(1+2e^{-3 \beta / 2T_s})^2}
+ {\it O} x^2 \right ]
\label{eq:T_s3}
\end{equation}

For $\beta=0$, the $x^2$ and higher terms vanish and
$T_s=3 \kappa (1-x)$ as expected for simple site
dilution \cite{Ziman89}.
The mean field theory overestimates the $x_c$ at
which $T_s$ vanishes because it does not contain the physics
of percolation.
As $\beta / T_s$ is increased, the coefficient
of $dT_s/dx$ increases; for
$\beta / T_s \rightarrow \infty$,
$T_s \rightarrow 3 \kappa (1-6x)$,
suggesting $x_c \approx 0.16$.
Comparison to the percolation
argument given previously suggests
that this is an underestimate.
The general result, however, of a $T_s(x)$
which drops rapidly as $x$ is increased
and depends somewhat on model parameters
(and so on materials), is in reasonable
accord with data.
Note however that at low T quantum effects
involving motion of holes will become important.
\section{Conclusion}
A classical model for $\lax$
has been analyzed.
It is known that doping on the La
site changes the valence of the Mn
site in such a way that the mean number
of outer d-shell electrons on the Mn is 1-x.
The holes were assumed to be classical, so that
each Mn site is occupied, with probability 1-x, or
empty,  with probability x.
The hypothesis of classical holes has been
shown to be consistent with the resistivity
at all $x$ and $T > 400K$ and for $x < x_{cond} \sim 0.3$
and all $T$ \cite{Millis95b}.

Because the outer Mn d-orbital is twofold degenerate,
a Jahn-Teller distortion of the surrounding
oxygen octahedron which lowers the local
cubic symmetry to tetragonal is may
occur about each occupied Mn site, while a
breathing mode distortion may occur
around each unoccupied site.
Each oxygen is shared by two Mn ions, so
distortions on adjacent sites are coupled.
The coupling was determined from a classical
harmonic approximation to the lattice dynamics.
The parameters of the model were determined
by fitting to the structural data obtained
for $LaMnO_3$.
The principal results are:

1) The basic energy gained in a local
Jahn-Teller distortion,
$E_{0} \gtrsim 0.1$ eV.  The estimate
$E_0 \approx 0.6$ eV was obtained using a phonon
frequency estimated from an optical measurement on
$La_{1.85}Sr_{.15}MnO_3$.  A direct measurement of
the splitting $4E_0$ between the two d-levels would
be desirable.
The distortions are in any event well
formed at any relevant temperature and
the structural transition is  to be
regarded as an order-disorder transition, at which local
Jahn-Teller distortions become spatially decorrelated,
but do not disappear.

2) The model describing the transition is given
in Eq. \ref{eq:E} and may be approximated either by an
antiferromagnetic x-y model with a modest three-fold
anisotropy by or a three-state Potts model with an antiferromagnetic
first neighbor interaction and a weak
second neighbor interaction.
Which model is more nearly correct depends on whether
the anharmonicity parameter A is larger or smaller than
the stiffness $\kappa$ which orients the distortions.
By combining an optical measurement of the highest phonon
frequency in $La_{1.85}Sr_{0.15}MnO_3$ with a calculation of
$T_s$ the estimate $A \sim \kappa$ was obtained.

3) Added holes disrupt the long range order
by producing an effectively random field, which mis-orients
nearby Jahn-Teller distortions.
It would be very interesting if it were possible to
observe directly this local misorientation.
This random field effect was shown by various mean-field
calculations to lead to a rapidly
decreasing $T_s(x)$, in qualitative accord with data.
A Monte-Carlo investigation of the problem would be useful.

The results in this paper substantiate to some
degree the proposal\cite{Millis95a,Millis95b}
electron-lattice interaction is so strong that
the high $T$ cubic (or pseudocubic)
$0.2 \lesssim x \lesssim 0.4$ phase
of $\lax$ should be modelled as a disordered
array of polarons.
The results presented here provide a basis for
calculating polaron binding energies and mobilities, both
for $0.2 \lesssim x \lesssim 0.4$ and high $T$
and for low $x$ at all $T$.
\subsection*{Acknowledgements}
I thank B. I. Shraiman for many helpful conversations,
D. A. Huse for discussions of random field problems,
and C. M. Varma for discussions of electronic
correlation effects.

\appendix
\section{Derivation of Energy}

This Appendix  outlines the derivation of
Eq. \ref{eq:E} from Eqs. \ref{eq:E_{Latt}},
\ref{eq:E_{JT}}, and \ref{eq:E_{hole}1} and
discusses anharmonic terms.
Define
\begin{equation}
u_i^a = \sum_k e^{-ik \cdot R_i}
u_k^a
\label{eq:u}
\end{equation}
and similarly $\delta_k^a$.
Now $\delta_k^a$ may be decoupled from
$E_{Latt}$, Eq. \ref{eq:E_{Latt}},
by defining
\begin{equation}
\bar{\delta}_k^a = \delta_k^a -
\frac{1}{2} \;
\frac{K_1 (1+e^{ik_a})}{2K_1+K_2-K_2 cos k_a}  \; u_k^a
\label{eq:bar{delta}}
\end{equation}
$\bar{\delta}_k^a =0$ gives the equilibrium positions
about which the Mn ions fluctuate.
After decoupling,
the relevant part of the lattice energy may be written
\begin{equation}
E_{Latt} = K_1 \sum_{ka}
F(k_a) u_k^a u_{-k}^a
\label{eq:E_{Latt-1}}
\end{equation}
with
\begin{equation}
F(k_a) = \frac{1}{2} \;
\frac{(K_1+2K_2)(1-cosk_a)}{K_1+K_2-K_2 cos k_a}
\label{eq:F}
\end{equation}
The interaction energies are most conveniently
written in terms of the variables
$c_k^a$ defined via
\begin{equation}
c_k^a= \sum_i e^{ik \cdot R_i}
(1-h_i) cos 2( \theta_i+\psi^a)
\label{eq:c}
\end{equation}
where $\psi^z=0$, $\psi^x= - \pi / 3$ and
$\psi^y= \pi /3$ were introduced in
Eq. \ref{eq:E}.
Combining Eqs.
\ref{eq:E_{JT}}, \ref{eq:v_i^a}, \ref{eq:c}, gives
\begin{equation}
\begin{array}{rl}
E_{JT} & = \lambda \sum_{ka}
(1-e^{-ik_a}) \\
\label{eq:E_{J-T-a}}
E_{hole} & = \beta \lambda \sum_{ka} h_k
(1-e^{-ik_a}) u_{-k}^a
\label{eq:E_{hole-1}}
\end{array}
\end{equation}
The displacement $u$ may be eliminated by
writing $E$ in terms of
\begin{equation}
\bar{u}_k^a = u_k^a -
\frac{\lambda (1-e^{-ik_a})}{2K_1F(k_a)}
(c_k^a + \beta h_k)
\label{eq:baru}
\end{equation}
Again $\bar{u}=0$ defines the average state
about which the oxygen
atoms fluctuate.
The electronic part of the energy may then be written
\begin{equation}
E=- \frac{1}{2} \frac{\lambda^2}{K_1}
\sum_{ka}
\frac{1-cosk_a}{F(k_a)}
(c_k^a + \beta h_k) (c_{-k}^a +
\beta h_{-k} )
\label{eq:E-1}
\end{equation}
Fourier transformation yields Eq. \ref{eq:E}
except for the term proportional to A.
This term arises from a lattice anharmonicity of
the form $\sum_i v_i^3$.
Use of Eq. \ref{eq:baru}
yields several terms, of which the largest is
$A cos (6 \theta)$.

\section{Analysis of Structure}

In this Appendix the $LaMnO_3$ structural data obtained
by Ellemans et. al. \cite{Ellemans71}
are analyzed.  The magnitudes of the atomic displacements
observed in $LaMnO_3$ in ref \cite{Ellemans71} are somewhat
greater than those reported in previous work \cite{Matsumoto70}.
Indeed, the displacements reported for $LaMnO_3$
by Ref \cite{Matsumoto70}
are very similar to those reported by Ellemans et. al.
for $La_{1.95}Ca_{0.05}MnO_3$ \cite{Ellemans71}.
It will be assumed here that the larger values are most representative
of the undoped material, and that the $LaMnO_3$ sample studied
in Ref \cite{Matsumoto70} was inadvertantly doped.

The actual crystal structure of $LaMnO_3$ is complicated;
for example every Mn-O-Mn bond is buckled.
It is assumed here that the important quantities
are the Mn-O bond lengths, and that the remaining
distortions are subsidiary, being driven by rotations
of the (distorted) $MnO_6$ octahedra required to
fit the rigid $MNO_6$ octahedra into a lattice
with lattice constants smaller than twice the Mn-O
distances.

$LaMnO_3$ was found to be orthorhombic, with three unequal Mn-O
distances, which are
\begin{equation}
\begin{array}{rlrl}
u_x & = 2.187 \AA \\
u_y & = 1.905 \AA \\
u_z & = 1.956 \AA
\end{array}
\end{equation}

Here $u_x , u_y$ are the distances most nearly
parallel to the nearest neighbor Mn-Mn bonds in
the basic (orthorhombic a-c) plane.
The x-y directions are about $45^{\circ}$ rotated
from the orthorhombic a-c axes.
$u_z$ is the distance most nearly parallel
to the orthorhombic b axis.

The z-oxygen is equidistant from the Mn
above and below it; the x and y oxygen ions are
not equidistant from the in-plane Mn; indeed, if
one moves from one Mn to its in-plane nearest neighbor,
the roles are reversed.
We therefore assume that the observed Mn-O
bond lengths have been obtained from an ideal
peroviskite structure with Mn-O distance
of $u^0 = (u_xu_yu_z)^{1/3} \cong 2.01 \AA$
by composing a uniform tetragonal distortion
$\Delta^u =-0.112 \AA (1,-1/2 -1/2)$ and a staggered
distortion $\Delta^s =.15 \AA (1, -1,0)$.
The mean lattice constant is $4 \AA$; thus $\Delta^u$
corresponds to a uniform strain components
$e^{xx} =e^{yy}=.014$,
$e^{zz} =- .028$, while the staggered
distortion is $u^s=.038(1,-1,0)$.

\section{Solution of Mean Field Equations}

In this Appendix details are given of the solution
of the mean field equations and of the energetics of
small deviations from the mean field solution.

\noindent
{\it Solution:}
assume a two sublattice solution with
$\theta = \theta_1$ on one sublattice and
$\theta = \theta_2$ on the other.
Take $A > 0$ without loss of generality and
choose units in which $3 \kappa =1$.
Write
\begin{equation}
\theta_{1,2} = \frac{2n_{1,2}+1}{6}
\pi + \delta_{1,2}
\label{eq:theta_{1,2}}
\end{equation}
with
\begin{equation}
- \pi / 6 \leq \delta_{1,2} \leq \pi / 6
\label{eq:bounds}
\end{equation}
{}From Eq. \ref{eq:E} one has
\begin{equation}
E_{mF} = cos
\left [ \frac{2(n_1-n_2) \pi}{3}
+ 2 \delta_1 - 2 \delta_2 \right ] -
\frac{a}{6} (cos 6 \delta_1 + cos 6 \delta_2)
\label{eq:E_{mF}}
\end{equation}
with $a = A/ \kappa > 0$.

Minimizing yields
\begin{equation}
2 sin
\left [ \frac{2(n_1-n_2) \pi}{3}
+ 2 \delta_1 - 2 \delta_2 \right ]
= a \; sin 6 \delta_1
\label{eq:delta_1}
\end{equation}
and
\begin{equation}
sin 6 \delta_1 =- sin 6 \delta_2
\label{eq:delta_2}
\end{equation}

Equation \ref{eq:delta_2} implies either
$\delta_1= -\delta_2$ or
$\delta_1=\delta_2 + (2n+1) \pi /6$;
the latter solution would imply that the anharmonicity
energy vanishes.  Such an extremum
cannot produce an absolute energy minimum.
If $\delta_2=-\delta_1$ then Eq. \ref{eq:delta_1}
may be solved.
Define
\begin{equation}
\delta= \frac{\pi}{12} -
\frac{1}{2} Arc sin
\left [ \sqrt{\frac{1+a^2}{4a^2}} -
\frac{1}{2a} \right ]
\label{eq:delta}
\end{equation}
Then $\delta_1=-\delta_2=\delta$ if
$n_1-n_2=1$ or $-2$ and
$\delta_1=-\delta_2=-\delta$ in
$n_1-n_2=-1$ or 2.
A different formula applies if
$n_1=n_2$; however, this case may be seen
not to lead to the global energy minimum because if
$n_1=n_2$ then the intersite term is positive unless
$\pi /8 \leq | \delta | \leq \pi / 6$ in
which case the anharmonicity term is positive.
The mean-field energy is thus minimized by any
of the six configurations with $n_1 \neq n_2$
and appropriate $\delta_1$ and $\delta_2$.

It is instructive to suppose that at all
but one of the sites the angles take values
minimizing $E_{mF}$ and to study the
energy function $E_0$ of the remaining angle.
Assume the isolated site is on the ``1''
sublattice and $n_1=0$.
Then
\begin{equation}
E_0=2 cos
\left [ 2 \theta - \frac{\pi}{3}
2 \delta \right ] + \frac{a}{3}
cos 6 \theta
\label{eq:E_0}
\end{equation}

For $a < a^* 4/3$,
Eq. \ref{eq:E_0}
has only one minimum, at the $\theta$
which satisfies the mean field equation.
For $a >a^*$ there are three minima.
For $a \gg a^*$ these occur at
\begin{equation}
\theta_n= \frac{(2n+1) \pi}{6}
+ \frac{2 sin 2n \pi / 3}{3a}
+ {\it O} \frac{1}{a^2}
\label{eq:theta}
\end{equation}
and correspond to energies
\begin{equation}
E_n= - \frac{a}{3} +
2 cos \frac{2n \pi}{3} \frac{2}{\sqrt {3} a}
sin \frac{2n \pi}{3}
\label{eq:E_n}
\end{equation}

This is the expected form of the energy
of a three-state Potts model with a first neighbor
``antiferromagnetic'' and second-neighbor
``ferromagnetic'' interactions.
A precise mapping is discussed in Appendix D.

\noindent
{\it Small deviations:}
assume that on every site the angle is close to one of
the two-sublattice solutions; thus, if
$a=1 , 2$
\begin{equation}
\theta_i = \theta_a + \psi_i
\label{eq:theta_{i,a}}
\end{equation}
with $\theta_a$ given by Eqs. \ref{eq:theta_{1,2}},
\ref{eq:delta}
according to whether $i$ is on sublattice 1 or 2 and
$\psi$ small.
Substituting Eq. \ref{eq:theta_{i,a}}
into Eq. \ref{eq:E} and expanding yields
\begin{equation}
E=E_{mF}+
\sum_k \omega_k^2 \psi_k \psi_{-k}
\label{eq:E-psi}
\end{equation}
The energy $\omega_k^{(n_1,n_1)}$ depends on the quantities
$n_1,n_2$ describing the possible ordered states.
There are three independent choices (($n_1,n_2) = (0,1)$,
($1,2$), ($0,2)$), each picks out a preferred
axis
$a=x,y,z$.
We have ($\gamma_k=1-(cosk_x+cosk_y+cosk_z)$)
\begin{equation}
\begin{array}{rl}
\omega_k^{(0,1)} = \omega_k^x & =
6 a cos 6 \delta +
4 cos \left ( \frac{\pi}{3}
- 4 \delta \right ) \gamma_k \\
    & - 4 \left [ cos k_x - \frac{1}{2}
(cosk_y + cosk_z ) \right ]
\label{eq:omega^x}
\end{array}
\end{equation}
Similarly $\omega^{1,2} = \omega^y$
and $\omega^{0,2}= \omega^z$.
Note that because of the relation of the
angles $\psi$ to the physical lattice distortions,
a nearly uniform variation of $\psi$ corresponds to a
nearly staggered variation of the physical
lattice distortions.  For physically relevant values of
a the gap is relatively large and the dispersion small.

\section{Derivation of Potts Model}

This Appendix gives the details of the
derivation of Eq. \ref{eq:E_{Potts}} from
Eq. \ref{eq:E}.
It is assumed $A$ is so large that only angles
near those minimizing the anharmonicity energy
$A cos 6 \theta_i$ are allowed.
Thus  write
\begin{equation}
\theta_i = \phi_i^{\alpha} + \delta_i
\label{eq:phi^a}
\end{equation}
with $\phi_i^{\alpha}$ one of
$\phi_x= 5 \pi / 6$,
$\phi_4= \pi / 6$
and $\phi_z = \pi / 2$ and $\delta_i$ a
small deviation.
Note that the Jahn-Teller distortion
corresponding to $\phi^a$ is
$u_a - \frac{1}{2} ( u_b + u_c)$.
Substituting Eq. \ref{eq:phi^a}
into Eq. \ref{eq:E} and expanding gives
\begin{equation}
\begin{array}{rl}
\frac{E}{3\kappa}  & = - \frac{Na}{3}
+ \frac{1}{3} \sum_{ia}
cos [ 2 \phi_i^{\alpha} +
2 \psi_c ] cos [ 2 \phi_{i+a}^{\alpha} +
2 \psi_a ] \\
   & + 6a \sum_i \bar{\delta}_i^2 \\
   & - \frac{8}{27} a \sum_{iab}
sin [ 2 \phi_i^{\alpha} + 2 \psi_a ] sin
[ 2 \phi_i^{\alpha} + 2 \psi_b ] \\
& cos [2 \psi_{i+a}^{\beta} + 2 \psi_c ]
cos [ 2 \phi_{i+b}^{\gamma} + 2 \psi_b ]
\label{eq:E-2}
\end{array}
\end{equation}

Note $N$ is the number of sites in the crystal,
$a = A / \kappa$ and
$\bar{\delta}_i = \delta_i - \delta_i^{min}$ with
$\partial E / \partial \delta_i^{min} =0$.
In the large $A$ limit the coefficient of the
$\bar{\delta}$ term is large, so fluctuations in
$\bar{\delta}$ may be neglected.

The energy may be more conveniently written
in a discrete notation.
Denote the state on site $i$ by the
continuous variable $\delta_i$ and a
$Q_i$ which indicates discrete quantity to
which of $\phi_x , \phi_y , \phi_z$ the
angle $\theta_i$ is nearest.
Choose $Q_i$ to be a three
component vector with a 1 in one place
and 0 in the other two.
$Q_i=(1,0,0)$ means $\theta_i$ is close to $\phi_x$,
$Q_i=(0,1,0)$ means $\theta_i$ is close to $\phi_y$ and
$Q_i=(0,0,1)$ means $\theta_i$ is close to $\phi_z$.
The interaction term of order is then a $3x3$ matrix
${\bf I}^a$, which depends on the direction
$\hat{a}$ of the band connecting the two sites.
One finds
\begin{equation}
{\bf I}^x = \frac{1}{3}
\left ( \begin{array}{rrr}
1 & -1/2 & -1/2 \\
-1/2 & 1/4 & 1/4 \\
-1/2 & 1/4 & 1/4 \end{array} \right )
\label{eq:I}
\end{equation}
${\bf I}^y , {\bf I}^z$ are obtained by permuting both
row and column in the obvious way.

The term of order $1/a$ is a rather complicated three
site interaction; however the important new
physics of this term is the coupling it
induces between sites on the same sublattice.
To determine this coupling it is convenient to
restrict attention to configurations (favored
by the order 1 term) in which adjacent sites are in different
Potts states, i.e. to terms in the order $1/a$ term in
Eq. \ref{eq:E-2} in which $\beta \neq \gamma$ or not.
If site $i$ is taken to be in state $\alpha =x$, then
the only non-constant terms are when $a=-b \neq \hat{x}$.
If $a= \pm \hat{y}$ then the energy is
$-4/9a$ if both sites are in the ``y''
state, $-1/9a$ if both are in the ``z'' state, and
$1/9a$ if the two are in different states, i.e. we
may write
\begin{equation}
E^{(2)} =- \frac{1}{9a} \sum_{ib} \vec{Q}_i
( \vec{J}_b \cdot \vec{Q}_{i+b}
\vec{Q}_{i+2b}
\label{eq:E^{(2)}}
\end{equation}
with $b= \pm x,y,z$ and
\begin{equation}
J_b^x = \left ( \begin{array}{crrr}
0  &  0  &  0 \\
0  &  4  &  -2 \\
0  &  -2 &  1  \end{array} \right )
\label{eq:J_b}
\end{equation}
etc.

The principle effect of $E^{(2)}$ is to lift the
degeneracies of the antiferromagnetic three state Potts
model; this effect may be mimicked by a simple second
neighbor ferromagnetic interaction with magnitude $J^{\prime}$
fixed e.g. by the requirement that it reproduce Eq. \ref{eq:E_n}.

\section{Mean Field $T_s$ with Holes}

In the presence of a concentration $x$ of holes,
one expects $T_s(x)=T_{co} (1- \alpha x )$.
In this Appendix $\alpha$ is derived using a mean field theory.
In leading order in $x$ one need only consider configurations in
which one of the six neighbors of the distinguished
site, say the one in the $b$ direction, has a hole.
{}From Eq. \ref{eq:E(theta)}
one has
\begin{equation}
\begin{array}{rl}
E_b(\theta ) &= 2 \kappa
cos (2 \theta +2 \psi_b )
[c \; cos 2 \psi_b -s \; sin 2 \psi_b + \beta ] \\
& + 4 \sum_{p = \pm 1}
cos (2 \theta +2 \psi_b+ \frac{2\pi_p}{3})
[c cos ( 2 \psi_b + \frac{2\pi_p}{3} )
-s \; sin (2 \psi_b + \frac{2\pi_p}{3}
\label{eq:E_b}
\end{array}
\end{equation}

Substituting Eq. \ref{eq:E_b} into
Eq. \ref{eq:P(theta)}, Eq. \ref{eq:Z},
expanding in $c$ and $s$, rearranging, and
discarding terms proportional to
$cos 2 \theta sin 2 \psi_b$ or
$sin 2 \theta cos 2 \psi_b$, which will not contribute to
averages of interest gives
\begin{equation}
\begin{array}{rl}
Z_b &=  \int_0^{\pi}
\frac{d\theta}{\pi}
e^{-A cos 6 \theta / T - \beta cos 2 \theta / T} \\
& x \left [ 1- \frac{4c \kappa}{T}
cos 2 \theta cos 2 \psi_b -
\frac{6c\kappa}{T} sin 2 \theta sin 2 \psi_b \right ]
\label{eq:Z_b}
\end{array}
\end{equation}
and, using also Eq. \ref{eq:T_s-1}
and results of Appendix C,
\begin{equation}
\begin{array}{rl}
1  & = \frac{3\kappa}{T_s} [1-6x] - \frac{1}{c}
\sum_b \frac{1}{Z_b} \int_0^{\pi}
\frac{d\theta}{2\pi}
e^{Acos 6 \theta /T - \beta cos 2 \theta / T} \\
 & \left [ 1- \frac{4c\kappa}{T_s}
cos 2 \theta cos 2 \psi_b -
\frac{6c\kappa}{T_s} sin \theta sin 2 \psi_b \right ] \\
   & \times cos ( 2 \theta - 2 \psi_b)
\label{eq:T_s-E}
\end{array}
\end{equation}

The integrals in Eqs. \ref{eq:Z_b},
\ref{eq:T_s-E} may be expressed in terms of
products of Bessel functions
$I_n(A/T) I_m( \beta / T)$.
The expressions become simple when $A \rightarrow 0$ or
$\infty$ or $\beta \rightarrow 0$ or $\infty$, and lead after
straightforward calculations to
Eqs. \ref{eq:T_s-1}, \ref{eq:T_s-2}.

\newpage
\begin{table}[h]
\caption{Values of parameters deduced by fitting mean field theory
to structural data.}
\begin{tabular}{cccccc} \hline
$K_2/K_1$ & $2\theta_1^{(0)}$ & $\lambda / K_1$ & $E_0/K_1 \times 10^3$ &
$\kappa / K_1 \times 10^4$ & $A/ \kappa$ \\
0 & 80.9 & .044 & 2.9 & 0 & 0.70 \\
0.1 & 79.2 & .045 & 2.8 & 0.85 & 0.87 \\
0.3 & 75.7 & .045 & 2.5 & 1.9 & 1.31 \\
0.5 & 72.3 & .046 & 2.4 & 2.6 & 1.92 \\
0.75 & 68.2 & .047 & 2.4 & 3.4 & 3.31 \\
1 & 64.4 & .049 & 2.4 & 4.0  & 6.93
\end{tabular}
\end{table}
\end{document}